\documentclass{emulateapj}

\usepackage{natbib}
\slugcomment{Submitted August 30, 2012, Accepted January 14, 2013}

\shorttitle{VLBA imaging of radio-quiet NLS1s}
\shortauthors{Doi et al.}

\begin{document}

\title{Very Long Baseline Array Imaging of Parsec-scale Radio Emissions\\ in Nearby Radio-quiet Narrow-line Seyfert 1 Galaxies}


\author{Akihiro Doi\altaffilmark{1,2},  Keiichi Asada\altaffilmark{3}, Kenta Fujisawa\altaffilmark{4}, Hiroshi Nagai\altaffilmark{5}, Yoshiaki Hagiwara\altaffilmark{5}, Kiyoaki Wajima\altaffilmark{6},\\ and Makoto Inoue\altaffilmark{3}}

\altaffiltext{1}{The Institute of Space and Astronautical Science, Japan Aerospace Exploration Agency, 3-1-1 Yoshinodai, Chuou-ku, Sagamihara, Kanagawa 252-5210, Japan}\email{akihiro.doi@vsop.isas.jaxa.jp}
\altaffiltext{2}{Department of Space and Astronautical Science, The Graduate University for Advanced Studies, 3-1-1 Yoshinodai, Chuou-ku, Sagamihara, Kanagawa 252-5210, Japan}
\altaffiltext{3}{Academia Sinica Institute of Astronomy and Astrophysics, P.O. Box 23-141, Taipei 10617, Taiwan}
\altaffiltext{4}{The Research Institute of Time Studies, Yamaguchi University, 1677-1 Yoshida, Yamaguchi, Yamaguchi 753-8511, Japan}
\altaffiltext{5}{National Astronomical Observatory, 2-21-1 Osawa, Mitaka, Tokyo 181-8588, Japan}
\altaffiltext{6}{Shanghai Astronomical Observatory, Chinese Academy of Sciences, Shanghai 200030, China}

\begin{abstract}
We conducted Very Long Baseline Array~(VLBA) observations of seven nearby narrow-line Seyfert~1~(NLS1) galaxies at 1.7~GHz~($\lambda$18cm) with milli-arcsecond resolution.  This is the first systematic very long baseline interferometry~(VLBI) study focusing on the central parsec-scale regions of radio-quiet NLS1s.  Five of the seven were detected at a brightness temperature of $\ga 5 \times 10^6$~K and contain radio cores with high brightness temperatures of $> 6 \times 10^7$~K, indicating a nonthermal process driven by jet-producing central engines as is observed in radio-loud NLS1s and other active galactic nucleus~(AGN) classes.  VLBA images of MRK~1239, MRK~705, and MRK~766 exhibit parsec-scale jets with clear linear structures.  A large portion of the radio power comes from diffuse emission components that are distributed within the nuclear regions ($\la300$~pc), which is a common characteristic throughout the observed NLS1s.  Jet kinetic powers limited by the Eddington limit may be insufficient to allow the jets escape to kiloparsec scales for these radio-quiet NLS1s with low-mass black holes of $\la10^7\ \mathrm{M_\sun}$.    
\end{abstract}



\section{INTRODUCTION}\label{section:introduction}

Narrow-line Seyfert~1 galaxies~(NLS1s) are identified by their optical properties: a flux ratio [\ion{O}{3}]/H$\beta$$<$3, and Balmer lines only slightly broader than forbidden lines (\citealt{Osterbrock_Pogge1985}), defined as FWHM(H$\beta$)$<$2000~km~s$^{-1}$ \citep{Goodrich1989}.  NLS1s make up $\sim20$\% of an optically selected active galactic nucleus~(AGN) sample from the Sloan Digital Sky Survey \citep[SDSS; e.g.,][]{Zhou_etal.2006}.  Recently, $\gamma$-ray emissions have been detected from several very radio-loud NLS1s by the Large Area Telescope onboard the {\it Fermi} Gamma-ray Space Telescope \citep{Abdo_etal.2009a,Abdo_etal.2009b,Foschini2011}; it has been argued that these NLS1s may represent a third class of $\gamma$-ray emitting AGNs along with blazars and radio galaxies.  The detected $\gamma$-rays strongly suggest the presence of blazar-like phenomena as a consequence of relativistic beaming effects due to a fast jet aligned close to our line of sight \citep{Blandford_Konigl1979}.  Moreover, the presence of highly relativistic jets in radio-loud NLS1s has been implied on the basis of radio flux variability observations \citep{Zhou_etal.2003,Doi_etal.2006,Doi_etal.2011,Yuan_etal.2008}, very-long baseline interferometry~(VLBI) imaging \citep{Doi_etal.2006,Doi_etal.2011,Giroletti_Panessa2009,D'Ammando_etal.2012}, and modeling of the spectral energy distributions \citep{Zhou_etal.2007,Yuan_etal.2008,Abdo_etal.2009b}.  Kiloparsec (kpc)-scale radio morphology is associated with several radio-loud NLS1s, including two $\gamma$-ray emitting NLS1s \citep{Anton_etal.2008,Gliozzi_etal.2010,Doi_etal.2012}.  However, these powerful jet phenomena are observed in only a small part of the NLS1 populations.  In fact, the fraction of radio-loud objects\footnote{The radio loudness $R$ is defined as the ratio of the 5~GHz radio flux density and optical {\it B}-band flux density.  A threshold of $R=10$ separating radio-loud and radio-quiet objects \citep{Kellermann_etal.1989} is frequently used.} in the NLS1 population is significantly low \citep{Zhou_Wang2002,Stepanian_etal.2003,Greene_etal.2006,Zhou_etal.2006,Yuan_etal.2008}, $\sim7$\% with $R>10$ and $\sim2.5$\% with $R>100$ \citep{Komossa_etal.2006}, compared with normal Seyfert galaxies and quasars \citep[$\sim10$\%--15\%; e.g.,][]{Ivezic_etal.2002}.  Similarly, kpc-scale radio emissions are less common in NLS1s than in broad-line AGNs \citep{Doi_etal.2012}.  The radio jet characteristics of a huge amount of the radio-quiet NLS1 population behind these extremes are little known.

NLS1s as a class are considered to have weak jet activities.  A radio-imaging study at arcsecond resolution using the Very Large Array (VLA) revealed that nearby (radio-quiet) NLS1s, in most cases, are unresolved radio sources [$\lesssim300$~parsec~(pc)] with modest radio luminosities ($10^{20}$--$10^{23}$~W~Hz$^{-1}$) and steep radio spectra ($\alpha<-0.5$, where $S_\nu \propto \nu^{\alpha}$, and $S_\nu$ is the flux density at the frequency $\nu$).  Thus, NLS1s are not remarkably different from normal Seyfert galaxies at radio wavelengths (\citealt{Ulvestad_etal.1995}, see also \citealt{Moran2000,Zuther_etal.2011}), although they exhibit many distinctive optical--X-ray properties.  
In addition to the unusually narrow Balmer lines, NLS1s exhibit several extreme properties such as strong permitted \ion{Fe}{2} emission lines \citep{Boroson_Green1992}, rapid X-ray variability \citep{Pounds_etal.1995,Leighly1999b}, steep soft X-ray spectra \citep{Wang_etal.1996,Boller_etal.1996,Leighly1999a}, and frequently observed blue-shifted line profiles \citep{Zamanov_etal.2002,Boroson2005,Leighly_Moore2004}.  It has been suggested that these properties are related to high-mass accretion rates close to the Eddington limit \citep{Boroson_Green1992,Brandt_Boller1998,Sulentic_etal.2000,Mineshige_etal.2000} on relatively low-mass black holes \citep[$\sim 10^5$--10$^{7.5}$~M$_\sun$;][]{Peterson_etal.2000,Hayashida2000,Grupe_Mathur2004,Zhou_etal.2006}.  
A high-accretion rate on a low-mass black hole supports the radio-quiescence of the NLS1 class, 
according to many studies using large AGN samples: a positive correlation between radio loudness and the black hole mass \citep{Laor2000,Lacy_etal.2001,Dunlop_etal.2003,McLure_Jarvis2004,Metcalf_Magliocchetti2006} and an anti-correlation between radio loudness and the accretion rate \citep{Ho2002,Lacy_etal.2001,Maccarone_etal.2003,Greene_etal.2006}.  
Similarly, \citet{Merloni_etal.2003} discovered ``a fundamental plane of black hole activity'' in three-dimensional space consisting of the radio luminosity, X-ray luminosity, and black hole mass (ranging from $10^1$ to $10^{10}$~M$_{\sun}$) for samples of X-ray binaries, our Galactic center, low-luminosity AGNs, Seyfert galaxies, and quasars.  
These tendencies in jet activity can be attributed to nonlinear dependences on the black hole mass and accretion rate \citep{Heinz&Sunyaev2003}.  
Thus, the weak radio emissions of NLS1s as a class can be understood in the framework established for other AGN classes.

However, the picture of a nonthermal jet origin for the observed radio emissions of NLS1s has not been fully established yet, particularly for the radio-quiet subclass, because of the difficulty of VLBI observations of weak radio sources.  Previously, a nonthermal jet origin for radio-quiet quasars was established by a series of VLBI detections \citep{Blundell&Beasley1998,Ulvestad_etal.2005}.  
The first VLBI study focusing on the NLS1 class at milli-arcsecond~(mas) resolutions examined the very {\it radio-loud} narrow-line quasar PMN~J0948+0022 using the Very Long Baseline Array (VLBA) \citep{Doi_etal.2006}.  A Doppler-beaming effect of highly relativistic jets is required to explain the observed extremely high brightness and rapid flux variability in the VLBA images.  Subsequent $\gamma$-ray detection by {\it Fermi}/LAT followed \citep{Abdo_etal.2009a}.  \citet{Doi_etal.2007} detected five additional radio-loud NLS1s using the Japanese VLBI Network and established the picture of a nonthermal process in the radio-loud NLS1 subclass.  \citet{Gu_Chen2010} reported compact radio structures for three of the five radio-loud NLS1s using VLBA imaging.  \citet{Doi_etal.2011} found evidence of pc-scale jet structures in all the five radio-loud NLS1s and pointed out (1)~the presence of relativistic nonthermal jets, which are essentially the same as those of other radio-loud AGN classes and (2)~the existence of both apparently radio-loud NLS1s owing to Doppler beaming and intrinsically radio-loud NLS1s.  Several $\gamma$-ray emitting NLS1s have been observed in survey programs of VLBI radio sources: SBS~0846+513 \citep{Kovalev_etal.2007,D'Ammando_etal.2012}, 1H~0323+342 \citep{Zhou_etal.2007}, and PKS~1502+036 (\citealt{Fey_Charlot2000}, see also \citealt{Dellacasa_etal.1998} for a study of this object as a compact radio galaxy candidate) and are now being monitored by VLBA in the MOJAVE\footnote{Monitoring of jets in AGNs with VLBA experiments \citep{Lister_etal.2009}} program.  
On the other hand, there have been only a handful of images of radio-quiet NLS1s: MRK~766, AKN~564 \citep{Lal_etal.2004}, NGC~5506 \citep{Middelberg_etal.2004}, NGC~4051 \citep{Giroletti_Panessa2009}, and MRK~1239 \citep{Orienti_Prieto2010} as part of a study of Seyfert galaxies (not focused on NLS1s).  A systematic VLBI study of radio-quiet NLS1s is crucial for understanding the nature of their radio-quiescent central engines with the lowest-mass black holes at the highest accretion rates in AGN classes.  Furthermore, radio-loud NLS1s should be discussed as extreme examples on the basis of more solid knowledge of radio-quiet objects that are the majority in the NLS1 population.

In the present paper, we report VLBI imaging of seven nearby radio-quiet NLS1s.  In Section~\ref{section:sample}, we describe the sample selection.  In Section~\ref{section:observation}, we describe the observations and imaging processes.  The observational results are presented in Section~\ref{section:result}.  We discuss the observed radio properties of the radio-quiet NLS1s in Section~\ref{section:discussion1}.  Finally, we summarize our studies in Section~\ref{section:summary}.  The observed properties of individual radio-quiet NLS1s are described in Appendix~\ref{section:individualobject}.  Throughout this paper, a $\Lambda$CDM cosmology with $H_0=71$~km~s$^{-1}$~Mpc$^{-1}$, $\Omega_\mathrm{M}=0.27$, and $\Omega_\mathrm{\Lambda}=0.73$ is adopted \citep{Komatsu_etal.2009}.

\defcitealias{2004AJ....127..156G}{1}
\defcitealias{Veron_etal.2001}{2}
\defcitealias{2010ApJS..189...15K}{3}
\defcitealias{2000ApJ...533..631K}{4}
\defcitealias{2007ApJ...654..799R}{5}
\defcitealias{2005MNRAS.364..208D}{6}
\defcitealias{2001A_A...377...52W}{7}
\defcitealias{Kukula_etal.1998}{8}
\defcitealias{Schmitt_etal.2001}{9}
\defcitealias{Ho_Ulvestad2001}{11}
\defcitealias{Ulvestad_etal.1995}{12}
\defcitealias{Ulvestad_Wilson1984}{13}

\begin{table*}
\caption{Nearby radio-quiet NLS1 sample for VLBA Observations.\label{table1}}
\begin{center}
\begin{tabular}{lcccccccccccc}
\hline
\hline
Name & $z$ & FWHM(H$\beta$) & [O~$_\mathrm{III}$]/H$\beta$ & $R_{4570}$ & Ref. & $\Gamma$ & $\log{M_\mathrm{BH}}$ & Ref. & $S_\mathrm{1.4GHz}^\mathrm{total}$ & $R_\mathrm{*}$ & $\Theta$ & Ref.  \\
 &  & (km~s$^{-1}$) &  &  &  &  & ($M_\sun$) &  & (mJy) &  & (pc) \\
(1) & (2) & (3) & (4) & (5) & (6) & (7) & (8) & (9) & (10) & (11) & (12) & (13) \\
\hline
\object{MRK 110} & 0.0353  & 1760 & 0.71 & 0.26 & \citetalias{2004AJ....127..156G} & 2.59  & 6.75 & \citetalias{2000ApJ...533..631K} & 10.1 & 1.8 & 2100 & \citetalias{Kukula_etal.1998} \\
\object{MRK 705} & 0.0292  & 1790 & 0.41 & 0.36 & \citetalias{Veron_etal.2001} & 2.53  & 6.72 & \citetalias{2007ApJ...654..799R} & 8.5\tablenotemark{c} & 2.1 &  $<55$  & \citetalias{Schmitt_etal.2001} \\
\object{MRK 1239} & 0.0199  & 1075 & 1.29 & 0.63 & \citetalias{Veron_etal.2001} & 4.06  & 5.89 & \citetalias{2007ApJ...654..799R} & 62.8 & 0.69 & 46 & 10 \\
\object{NGC 4051} & 0.0023  & 1120 & 0.55 & 0.25 & \citetalias{Veron_etal.2001} & 3.04  & 6.13 & \citetalias{2000ApJ...533..631K} & 14.7\tablenotemark{d} & 0.62 & 330 & \citetalias{Ho_Ulvestad2001} \\
\object{MRK 766} & 0.0129  & 1630 & 0.35 & 1.83 & \citetalias{Veron_etal.2001} & 2.63  & 6.80 & \citetalias{2005MNRAS.364..208D} & 39.7 & 0.63 & 65 & \citetalias{Ulvestad_etal.1995} \\
\object{MRK 783} & 0.0672  & 1655 & 2.29 & 0.49 & \citetalias{Veron_etal.2001},\citetalias{2010ApJS..189...15K} & 1.30\tablenotemark{a} & 7.16 & \citetalias{2001A_A...377...52W} & 33.2 & 2.0 &  $<380$  & \citetalias{Ulvestad_Wilson1984} \\
\object{AKN 564} & 0.0247  & 865 & 0.92 & 0.67 & \citetalias{Veron_etal.2001} & 3.4\tablenotemark{b} & 6.75 & \citetalias{2000ApJ...533..631K} & 29.1 & 0.44 & 320 & \citetalias{Schmitt_etal.2001} \\
\hline
\end{tabular}
\end{center}

\tablecomments{
Col.~(1) source name; 
Col.~(2) redshift; 
Col.~(3) FWHM of H$\beta$ line width of broad-line component; 
Col.~(4) flux ratio of [\ion{O}{3}]$\lambda5007$ to H$\beta$; 
Col.~(5) flux ratio of \ion{Fe}{2} multiplets in the range 4434--4684$\mathrm{\AA}$ to H$\beta$; 
Col.~(6) reference for optical-line emissions; 
Col.~(7) {\it ROSAT} soft X-ray photon index at 0.1--2.4~keV \citep{Pfefferkorn_etal.2001}; 
Col.~(8) black hole mass; 
Col.~(9) reference for the black hole mass; 
Col.~(10) NRAO VLA Sky Survey (NVSS; $\sim45\arcsec$) flux density at 1.4~GHz as total flux density; 
Col.~(11) radio loudness defined as the ratio of $k$-corrected 5~GHz radio to optical $B$-band flux densities \citep{Kellermann_etal.1989}.  We calculated the radio loudness using the NVSS flux density and SDSS 7DR PSF magnitude \citep{2009ApJS..182..543A} extinction-corrected by assuming a Balmer decrement of $\mathrm{H\alpha/H\beta=3.1}$ for all sources, except AKN~564 for which we used a $V$-band nuclear magnitude \citep{2008MNRAS.385...53M}.  We assume $\alpha=-0.7$ for radio (except for MRK~1239, MRK~766, and AKN~564; see Appendix~\ref{section:individualobject}) and $\alpha=-0.61$ for optical \citep{2010ApJS..187...64G} for the $k$~corrections.  
Col.~(12) maximum extent of radio structure from nucleus in VLA images at arcsecond resolution; 
Col.~(13) reference for the radio size. 
}
\tablerefs{
(1)~\citealt{2004AJ....127..156G}, 
(2)~\citealt{Veron_etal.2001}, 
(3)~\citealt{2010ApJS..189...15K}, 
(4)~\citealt{2000ApJ...533..631K}, 
(5)~\citealt{2007ApJ...654..799R}, 
(6)~\citealt{2005MNRAS.364..208D}, 
(7)~\citealt{2001A_A...377...52W}, 
(8)~\citealt{Kukula_etal.1998}, 
(9)~\citealt{Schmitt_etal.2001}, 
(10)~this paper (Appendix~\ref{section:individualobject}), 
(11)~\citealt{Ho_Ulvestad2001}, 
(12)~\citealt{Ulvestad_etal.1995}, 
(13)~\citealt{Ulvestad_Wilson1984}. 
}
\tablenotetext{a}{A large uncertainty ($\pm0.72$)}
\tablenotetext{b}{\citealt{Boller_etal.1996}}
\tablenotetext{c}{Flux density from FIRST ($\sim5\arcsec$) at 1.4~GHz to avoid the contamination from an unrelated source $\sim30\arcsec$ south \citep{2005A_A...442..909G}}
\tablenotetext{d}{Flux density from VLA A-configuration ($\sim1\farcs2$) at 1.4~GHz \citep{Ho_Ulvestad2001}, because the NVSS emission is contaminated by the contribution of a host galaxy \citep{Baum_etal.1993}}
\end{table*}

\section{Sample}\label{section:sample}
A nearby radio-quiet NLS1 sample consisting of seven objects was made for this VLBI imaging study (Table~\ref{table1}).  
We selected from the 205~NLS1s listed in ``Quasars and Active Galactic Nuclei,'' 10th~Ed.~\citep{Veron&Veron2001} and the 64~NLS1s in ``Spectrophotometric Atlas of NLS1s'' \citep{Veron_etal.2001}.  We processed the following selection criteria step-by-step: (1)~redshift $z<0.1$ (129 sources survived), (2)~inclusion in the VLA Faint Images of the Radio Sky at Twenty-Centimeters \citep[FIRST;][]{Becker_etal.1995}, 03Apr11 version (60~sources survived), (3)~existence of a radio counterpart with an offset of $<$1\arcsec~in the FIRST image (33~sources survived), (4)~a peak intensity of more than 5~mJy~beam$^{-1}$ (13~sources survived), and (5)~availability of a radio position measured at a resolution of $<0\farcs5$~from literature or archival data (to enable processing by VLBA correlator as described in detail in Section~\ref{section:skyposition}; 7~sources survived).  
One of the selected NLS1s, MRK~1388, has sometimes been classified as a type~1.9 or 2 Seyfert \citep[e.g.,][]{Goodrich1989} with an unresolvable broad-line component~\citep{Osterbrock1985,Veron_etal.1997}; hence, we excluded it from our sample.  
In addition, we included AKN~564, one of the best-studied NLS1s, although it does not properly satisfy criterion~(2); we substituted VLA archival data~(AT149) obtained at $\sim5$~arcsec resolution as equivalent to the FIRST data.  
As a result, seven sources were selected.  They are genuine NLS1s according to the optical definitions and also other aspects such as the \ion{Fe}{2} line strength, soft X-ray spectrum, and black hole mass [Columns~(3)--(9) in Table~\ref{table1}].  All seven NLS1s are radio quiet by the definition of radio loudness using the nuclear optical and total radio flux densities [$R_*$; Column~(11) in Table~\ref{table1}].  They all, except for MRK~110, are compact radio sources covering $\la300$~pc in VLA images [Column~(12) in Table~\ref{table1}].

\section{OBSERVATIONS AND DATA REDUCTION}\label{section:observation}

\subsection{Sky Positions}\label{section:skyposition} 
The field-of-view~(FOV) in this study is limited for statistical reasons.  Searching for emission with a detection limit of $4\sigma$ requires a FOV limited to $\sim 32,000$ points, corresponding to a radius of $\sim100$~times the synthesized beam on the sky.  The expected size of the synthesized beam is about 11~mas$\times$5~mas aligned nearly north--south at the declinations of our sample: the FOV\footnote{The bandwidth smearing for a nominal continuum mode and the time-average smearing for 2-sec integration in the VLBA correlator restrict the FOV to $\sim7$\arcsec\ and $\sim15$\arcsec, respectively \citep{Thompson_etal.1986,Wrobel1995}.} is supposed to be $\sim$1\farcs1$\times$0\farcs5.  Hence, we attempted to determine the positions for correlation~(the phase-tracking centers) to sub-arcsec accuracy during sample selection (Section~\ref{section:sample}).  

For each NLS1, we obtained the radio position with an uncertainty of $\sim0\farcs1$ in images from the VLA A-array configuration in the archive (AM384, AK394, AB973, and AL410) and/or in the literature \citep{Ulvestad_Wilson1984,Ulvestad_etal.1995,Kukula_etal.1998,Nagar_etal.1999,Thean_etal.2000,Schmitt_etal.2001}.  We used the National Radio Astronomy Observatory~(NRAO) {\tt SCHED} software when it was necessary to convert a B1950 VLA position to J2000.  The phase-tracking centers used in our VLBI observations are listed in Table~\ref{table2}.

\subsection{Observations}\label{section:VLBAobservation}
We observed the seven NLS1s on May~09, 2005 at 1.7~GHz using ten antennas of the VLBA at the NRAO (observation code BD106).   
A left circular polarization was received at a center frequency of 1.667~GHz with a total bandwidth of 32~MHz.  We observed in phase-referencing mode, which allows us to derive calibration parameters for instrumental and atmospheric effects from the observations of nearby strong compact radio sources \citep[``calibrators;''][]{Beasley&Conway1995}.  The phase-reference cycle periods were 5~min, and the total on-source time was about 40~min for each target.

\begin{table*}
\caption{Parameters of phase-reference VLBA observations.\label{table2}}
\begin{center}
\begin{tabular}{lccccccc}
\hline
\hline
      Name & \multicolumn{ 2}{c}{Phase-tracking center (J2000)} &       Ref. & \multicolumn{ 2}{c}{Astrometric position (J2000.0)} &  Reference & $\Delta\theta$ \\

           &       R.A. ($^\mathrm{h}$ $^\mathrm{m}$ $^\mathrm{s}$) &      Decl. (\degr\ \arcmin\ \arcsec) &            &       R.A. ($^\mathrm{h}$ $^\mathrm{m}$ $^\mathrm{s}$) &      Decl. (\degr\ \arcmin\ \arcsec) &     Source &      (\degr) \\

       (1) &        (2) &        (3) &        (4) &        (5) &        (6) &        (7) &        (8) \\
\hline
   MRK 110 & 09 25 12.8500 & $+$52 17 10.388 &         1  & 09 25 12.8476 & $+$52 17 10.387 & J0932+5306 &       1.40 \\

   MRK 705 & 09 26 03.2763 & $+$12 44 03.659 &         2  & \ldots\tablenotemark{a} &     \ldots & J0931+1414 &       1.94 \\

  MRK 1239 & 09 52 19.0996 & $-$01 36 43.553 &         3  & 09 52 19.0980 & $-$01 36 43.546 & J0948+0022 &       2.16 \\

  NGC 4051 & 12 03 09.6105 & $+$44 31 52.680 &         4  & \ldots\tablenotemark{b} &     \ldots & J1155+4555 &       1.98 \\

   MRK 766 & 12 18 26.5186 & $+$29 48 46.471 &         5  & \ldots\tablenotemark{a} &     \ldots & J1217+3007 &       0.33 \\

   MRK 783 & 13 02 58.8430 & $+$16 24 27.748 &         6  & 13 02 58.8438 & $+$16 24 27.777 & J1311+1658 &       2.09 \\

   AKN 564 & 22 42 39.3397 & $+$29 43 30.915 &         2  & \ldots\tablenotemark{b} &     \ldots & J2236+2828 &       1.85 \\
\hline
\end{tabular}

\tablecomments{Col.~(1) source name; Col.~(2)--(3) position of phase-tracking center used in VLBA correlator; Col.~(4) reference used to determine the phase tracking center; Col.~(5)--(6) astrometric positions determined by our VLBA observations; Col.~(7) reference calibrator; Col.~(8) separation angle between target and reference calibrator.}

\tablerefs{(1)~average of the results of \citealt{Ulvestad_Wilson1984}, \citealt{Kukula_etal.1998}, and VLA archival data AL410; (2)~AB973; (3)~average of the results of \citealt{Thean_etal.2000}, \citealt{Ulvestad_etal.1995}, AK394, and AM384; (4)~\citealt{Hagiwara_etal.2003}; (5)~average of the results of \citealt{Kukula_etal.1995} and \citealt{Thean_etal.2001}; (6)~\citealt{Ulvestad_Wilson1984}.}

\tablenotetext{a}{Position is not available because of detection in self-calibrated image}
\tablenotetext{b}{Non detection}
\end{center}
\end{table*}

\subsection{Data Reduction}\label{section:datareduction}
Standard calibration procedures for VLBA phase referencing were applied during data reduction using the Astronomical Image Processing System \citep[{\tt AIPS};][]{Greisen2003}.  Inappropriate earth-orientation parameters were corrected using the {\tt AIPS} task {\tt CLCOR}.  Amplitude calibration using a-priori gain values together with system temperatures measured during the observations was applied; the accuracy is typically 5\%.  The ionospheric dispersive delay was corrected using a global ionospheric model and the task {\tt TECOR}.  The time and spectral domains were carefully inspected and data flagging was performed; there was no significant radio interference in the data.

Conservatively, we first searched for emission components with peak intensities of $>5\sigma$ of the noise on images with natural weighting in a FOV of $1\farcs1 \times 0\farcs5$, and also greater than $6\sigma$ in a FOV of $3\farcs0 \times 1\farcs3$ (Section~\ref{section:skyposition}).  At this stage, we detected three sources: MRK~110, MRK~1239, and MRK~783.  They initially appeared at positions offset by less than several tens of mas from the phase-tracking centers.  After correcting the map centers to the emission peaks, we deconvolved the images using the task {\tt IMAGR} (CLEAN) and measured the astrometric positions of the emissions using the task {\tt JMFIT}~(Table~\ref{table2}).  The images of negative detection sources MRK~705, MRK~766, and AKN~564 showed quadrangular patterns resembling beam patterns.  There were several emission candidates with intensities of 4$\sigma$--5$\sigma$ in the images.  The patterns seemed to contribute to increasing the image noise to levels significantly higher than the expected thermal noises.  This suggested that significant emission does exist, but its power could be scattered owing to imperfect calibration of the phase-referenced array.  The quality of the phase referencing is usually more or less degraded by the uncertainty in the correlator models, mainly regarding the residual atmospheric (troposphere and ionosphere) delay for each antenna's site.  Hence, we attempted self-calibration in phase with a signal-to-noise ratio threshold of $2.5$ using the task {\tt CALIB}.  Astrometric information was lost in the self calibration procedure.  We defined the success of self-calibration according to whether the solutions were found as a smooth-phase gradient for all the antennas through more than half the observing time on a target.  The self-calibration was successful and emissions were detected for MRK~705, MRK~1239, MRK~766, and MRK~783.  We smoothed and interpolated the solutions.  The deconvolution and self-calibration algorithms were interactively applied a few times.  
As a result, NGC~4051 and AKN~564 were not detected with a detection limit of 5~times the image noise.  

We used the {\tt Difmap} software \citep{Shepherd1997} to make final images using the CLEAN algorithm from the calibrated visibilities (additional self-calibration was not performed in {\tt Difmap}).  We used step-by-step uniform, natural, and ($u$,~$v$)-tapered weighting functions for CLEAN to retrieve both compact and diffuse components as much as possible.  Consequently, we obtained self-calibrated images for four targets~(MRK~705, MRK~1239, MRK~766, and MRK~783) and a direct image for one target~(MRK~110).

\section{RESULTS}\label{section:result}
\begin{figure*}
\epsscale{0.8}
\plotone{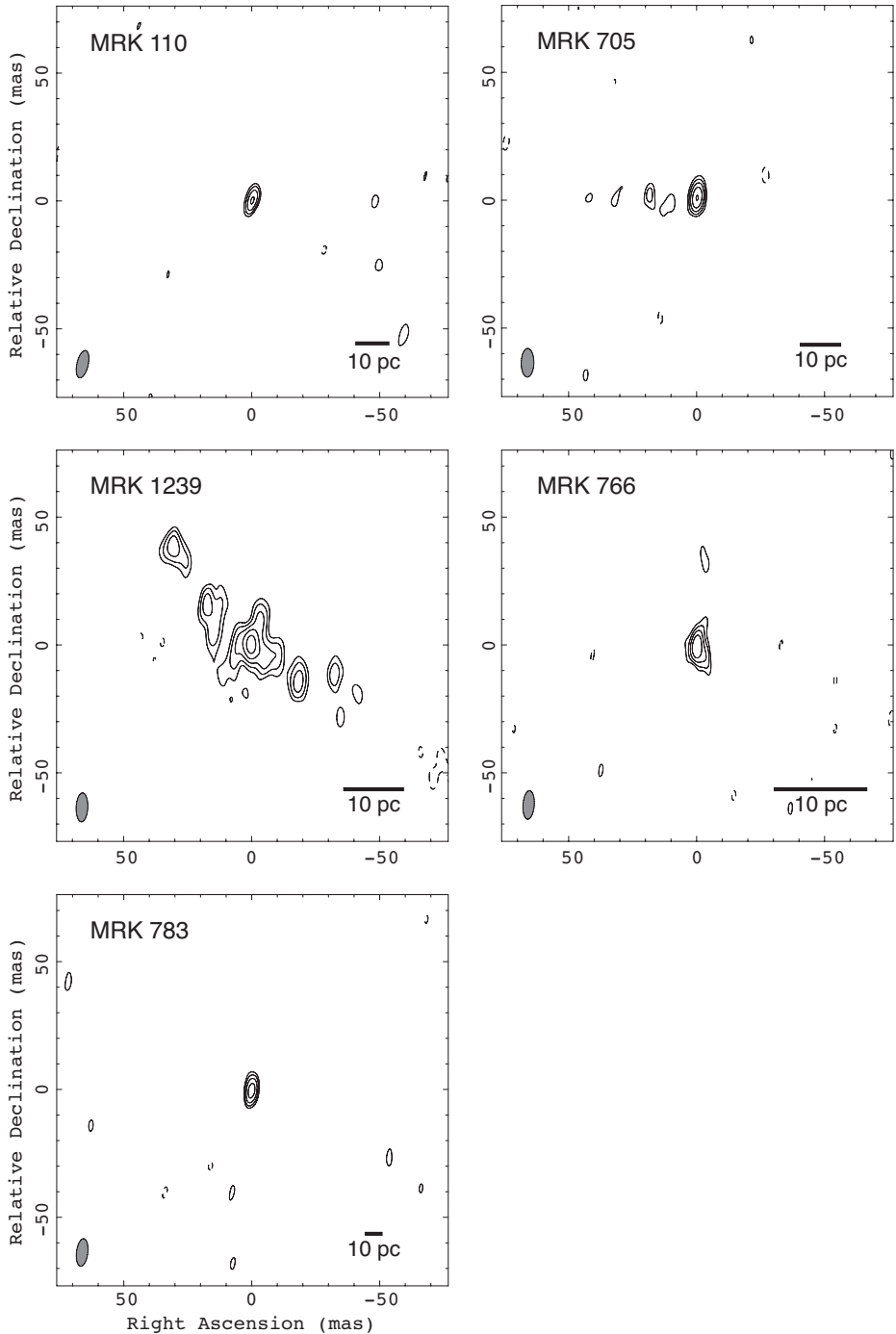}
\figcaption{VLBA images with natural weighting for detected nearby radio-quiet NLS1s at 1.7~GHz in the region of $150 \times 150$~mas$^{2}$.  Contour levels are separated by factors of $\sqrt{2}$ beginning at 3 times the rms noise~(Table~\ref{table3}).  Negative and positive contours are represented by dashed and solid curves, respectively.  Half-power beam sizes are given in the lower left corners.  Angular scales corresponding to a linear scale of 10~pc are given in the lower right corners. 
\label{figure1}}
\end{figure*}

Five of the seven radio-quiet NLS1s were detected in the VLBA observations.  The results of the image analyses are shown in Table~\ref{table3}.  The images with natural weighting for the detected sources are shown in Figure~\ref{figure1}.  We also show images convolved with a restored beam size of 10~pc in Figure~\ref{figure2}.  We describe the details for individual radio-quiet NLS1s in Appendix~\ref{section:individualobject}.  

\subsection{Brightness Temperatures}\label{section:TB}
The images of the five detected radio-quiet NLS1s show a compact component (core) with high brightness temperatures of $T_\mathrm{B}>6\times10^7$~K (Table~\ref{table3}), which were calculated from 
\begin{equation}
T_\mathrm{B} = 1.8 \times 10^9 (1+z) \frac{S_\nu}{\nu^2 \phi_\mathrm{maj}\phi_\mathrm{min}} 
\label{equation:TB}
\end{equation}
in Kelvin at the rest frame, where $z$ is the redshift, $S_\nu$ is the flux density in mJy at the observing frequency $\nu$ in GHz, and $\phi_\mathrm{maj}$ and $\phi_\mathrm{min}$ in mas are the fitted full widths at half-maximum of the major and minor axes of the component, respectively \citep[cf.][]{Ulvestad_etal.2005}.  If a source was unresolved, we adopted half the beam sizes as the upper limits to the source sizes ($\phi<\theta/2$).  

The typical image noise of $\sim0.41$~mJy~beam$^{-1}$ corresponds to $T_\mathrm{B}\sim1.7\times10^6$~K.  Thus, VLBI imaging works as a high-brightness filter.  Even diffuse emissions of $\sim3\sigma$ (the lowest contour level in the images) imply $T_\mathrm{B} \approx 5\times10^6$~K, which is too high to attribute to any stellar origin.  We conclude that all the detected VLBA emissions from these radio-quiet NLS1s are relevant to the activity of AGNs.

\begin{figure*}
\epsscale{0.8}
\plotone{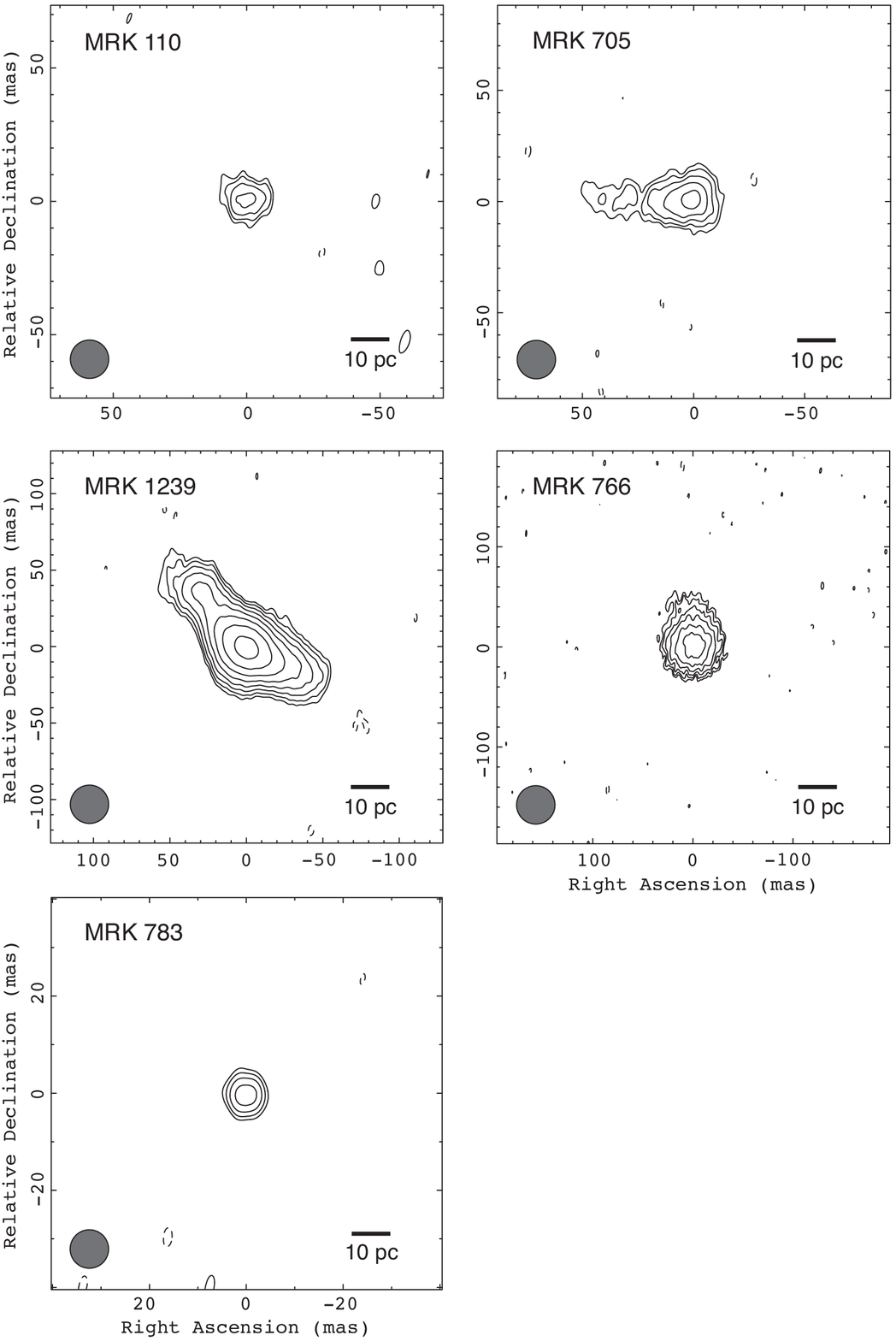}
\figcaption{VLBA images restored with the 10-pc beam in the regions of $205 \times 250$~pc$^{2}$.  Map descriptions are the same as for Figure~\ref{figure1}. 
\label{figure2}}
\end{figure*}

\subsection{Source Structures}\label{section:sourcestructure}
The five detected sources have compact emissions with high brightness temperatures.  Furthermore, we can recognize significant structures in the images of MRK~1239 (Figure~\ref{figure1}), MRK~705, and MRK~766 (Figure~\ref{figure2}).  A linear structure that extending to either side of a brightness peak was clearly seen in MRK~1239.  The visible radio extent was $\sim$100~mas (corresponding to $\sim40$~pc), which was consistent with a slight elongation found in a VLA archival image at sub-arcsec resolution [Figure~\ref{figureA} (b) in Appendix~\ref{section:individualobject}].  MRK~705 showed a one-sided linear structure $\sim45$~mas~($\sim26$~pc) east of the core.  No larger structure has been found in VLA images (Appendix~\ref{section:individualobject}).  MRK~766 showed a signature of a northern elongation of $\sim50$~mas ($\sim13$~pc) in the restored image (Figure~\ref{figure2}).   The VLBA images for the other two sources (MRK~110 and MRK~783) exhibit apparently compact emission.

\begin{table*}
\caption{Results of VLBA observations.\label{table3}}
\begin{center}
\begin{tabular}{lccccccccccc}
\hline
\hline
Name & $I_\mathrm{1.7GHz}^\mathrm{VLBIcore}$ & $S_\mathrm{1.7GHz}^\mathrm{VLBIcore}$ & $S_\mathrm{1.7GHz}^\mathrm{VLBItotal}$ & $\sigma$ & $\theta_\mathrm{maj}$ & $\theta_\mathrm{min}$ & P.A. & $T_\mathrm{B}$ & $\theta_\mathrm{10pc}$ & $S_\mathrm{1.7GHz}^\mathrm{<10pc}$ & $f_\mathrm{1.7GHz}^\mathrm{10pc}$ \\
 & (mJy beam$^{-1}$) & (mJy) & (mJy) & (mJy beam$^{-1}$) & (mas) & (mas) & (\degr) & ($\times10^7$~K) & (mas) & (mJy) \\
(1) & (2) & (3) & (4) & (5) & (6) & (7) & (8) & (9) & (10) & (11) & (12) \\
\hline
MRK 110 & $1.1\pm0.1$ & $1.2\pm0.2$ & 1.1 & 0.128 & 10.9 & 4.4 & $-$12.7 & $>6.5$ & 14.4 & 1.1 & 0.13 \\
MRK 705 & $1.8\pm0.2$ & $1.8\pm0.3$ & 3.3 & 0.144 & 11.2 & 4.9 & $-$0.8 & $>8.7$ & 17.3 & 1.8 & 0.24 \\
MRK 1239 & $2.4\pm0.2$ & $4.8\pm0.5$ & 20.0 & 0.169 & 11.3 & 4.5 & $-$2.5 & 7.4  & 25.1 & 13.1 & 0.23 \\
NGC 4051 & $<0.7$ & $<0.7$ & $<0.7$ & 0.143 & 11.2 & 4.3 & \ldots & $<1.0$ & 121 & $<0.7$ & $<0.05$ \\
MRK 766 & $1.3\pm0.1$ & $1.6\pm0.2$ & 2.5 & 0.103 & 11.2 & 4.4 & $-$4.1 & 8.4  & 38.4 & 2.4 & 0.07 \\
MRK 783 & $1.4\pm0.1$ & $1.3\pm0.2$ & 1.3 & 0.129 & 10.8 & 4.4 & $-$6.7 & $>7.7$ & 7.9 & 1.3 & 0.04 \\
AKN 564 & $<1.0$ & $<1.0$ & $<1.0$ & 0.208 & 11.8 & 4.5 & \ldots & $<1.3$ & 20.4 & $<1.0$ & $<0.04$ \\
\hline
\end{tabular}
\end{center}
\tablecomments{Col.~(1) source name; Col.~(2) peak intensity of core component; Col.~(3) flux density of core component, measured by a single Gaussian profile fitting using the {\tt AIPS} task {\tt JMFIT}; Col.~(4) total flux density retrieved by CLEAN components; Col.~(5) image rms noise on blank sky; Col.~(6)--(8) synthesized beam pattern in major axis, minor axis, and position angle, respectively; Col.~(9) brightness temperature of core component, calculated using Equation~(\ref{equation:TB}); Col.~(10) angular scale in mas corresponding to a linear scale of 10~pc; Col.~(11) flux density from central 10~pc at 1.7~GHz in the observer frame; Col.~(12) radio core fraction, defined as the ratio of the flux density from the central 10~pc to the total flux density at 1.7~GHz in the rest frame.  The extended flux component was estimated by subtracting the contribution from the central 10~pc [Col.~(11)] from the NVSS total flux density at 1.4~GHz (Table~\ref{table1}), assuming the core's spectral index available in the literature or $\alpha=-0.5$ for values unavailable in the literature and $\alpha=-0.7$ for the extended component.  See Appendix~\ref{section:individualobject} for individual objects.}
\end{table*}

The VLBA-detected flux density was significantly lower than the VLA total flux density for all the sources, indicating low fractions of high-brightness components ($\ga5\times10^6$~K) relative to lower-brightness components (Section~\ref{section:TB}).  In this paper, the radio core fraction, $f_\mathrm{1.7GHz}^\mathrm{10pc}$, is defined as the ratio of the flux density from the central 10~pc (in diameter) to the total flux density at 1.7~GHz in the rest frame.  At the distances to our samples, 10~pc typically corresponds to $\sim20$~mas [Column~(10) in Table~\ref{table3}].   
Our VLBA baselines can retrieve $\geq95$\% of the emissions as a correlated flux density at the minimum (projected) length of $\sim210$~km for the 20-mas radio size.  
Hence, the fluxes from the central 10~pc were practically measurable [Column~(11) in Table~\ref{table3}].  
On the other hand, we cannot rule out the possibility that emissions with much larger angular scales were partially or scarcely detected (``resolved out'').  Hence, we adopted the total flux densities from the NRAO VLA Sky Survey \citep[NVSS;][]{Condon_etal.1998} at 1.4~GHz as listed in Table~\ref{table1}; 
the contribution of extended component was calculated by subtracting the $<10$~pc contribution at 1.4~GHz from the NVSS value at the rest frame, and then, converted to the value at 1.7~GHz, assuming spectral indices of compact and extended components (see, Appendix~\ref{section:individualobject} and Table~\ref{table3} in detail).     
Consequently, the core fractions were typically less than 10\%, up to 24\% at most [Column~(12) in Table~\ref{table3}].  Because the entire radio structures are $\la300$~pc in the VLA images [Column~(12) in Table~\ref{table1}], a significant portion of the emissions should originate in the regions of $\sim10$--300~pc in the form of low-brightness components detected ($\ga 5 \times 10^6$~K) and not detected ($\la 5 \times 10^6$~K) in the VLBA images.  
In fact, we recognized a clear enhancement in the visibility amplitude of MRK~1239 at baselines of $\la3$~M$\lambda$ [Figure~\ref{figureA} (a) in Appendix~\ref{section:individualobject}] and marginal enhancements also in MRK~766 and AKN~564 (Appendix~\ref{section:individualobject}).  These are evidences for the presence of a significant amount of emission components with sizes of $>70$~mas, corresponding to $>20$--30~pc for these sources.

\section{DISCUSSION}\label{section:discussion1}
\subsection{Compact Emission Components}\label{section:compactemission}
Our VLBA imaging provides the first insight into the pc-scale region of several nearby radio-quiet NLS1s in a radio band.  
All five detected radio-quiet NLS1s have a compact radio component with a high-brightness temperatures of $>6\times10^7$~K.  This is an evidence of {\it a nonthermal process driven by jet-producing central engines in radio-quiet NLS1s}, as well as radio-quiet Seyfert nuclei \citep[e.g.,][]{Preuss&Fosbury1983,Neff&Bruyn1983} and radio-quiet quasars \citep{Blundell&Beasley1998,Ulvestad_etal.2005}.  

It is notable that the central components make only minor contributions compared to the VLA total flux densities (i.e., low values of $f_\mathrm{1.7GHz}^\mathrm{10pc}$; Section~\ref{section:sourcestructure}).  A previous study using the VLBA reported that all but 1 of 39~radio-selected low-luminosity AGNs were detected, and they exhibited core-dominated structures in most cases (\citealt{Nagar_etal.2005}, see also \citealt{Falcke_etal.2000,Anderson_etal.2004}).  However, the radio-quiet NLS1s in our sample present a contrasting aspect, although both of the AGN subclasses are weak radio sources in the nearby universe.  The radio spectra are also quite different: flat or mildly inverted spectra in low-luminosity AGNs \citep{Nagar_etal.2001,Anderson_etal.2004,Doi_etal.2005,Doi_etal.2011,Doi_etal.2013} but steep spectra in NLS1s \citep{Moran2000,Zhou_Wang2002}.  Such flat or inverted spectra would be accounted for by optically thick synchrotron emitters at the base of the jets \citep[e.g.,][]{Falcke&Biermann1999}.  The scale-invariant jet model predicts that the radio luminosity from the jet bases depends on $\propto M_\mathrm{BH}^{17/12}$ in most accretion scenarios, where $M_\mathrm{BH}$ is the black hole mass \citep{Heinz&Sunyaev2003}.
In fact, the known radio sources of low-luminosity AGNs tend to have large black holes of $M_\mathrm{BH}\sim10^8$--$10^9$~M$_\sun$ \citep[e.g.,][]{Nager_etal.2002b} in contrast to values of $\sim 10^6$--$10^7$~M$_\sun$ for the radio-quiet NLS1s in our sample (Table~\ref{table1}) and the majority of nearby NLS1s \citep[e.g.,][]{Doi_etal.2012,Grupe_Mathur2004,Zhou_etal.2006}.  Hence, compared with low-luminosity AGNs, the radio-quiet NLS1s possess less efficient radio cores.

Moreover, the radio-quiet NLS1s also exhibit significantly lower core contributions than radio-loud NLS1s \citep{Doi_etal.2011}; core enhancement due to the Doppler beaming effect is a natural explanation in some cases \citep{Doi_etal.2007,Gu_Chen2010}.  In our VLBA images with limited dynamic ranges, we found no clear evidence of highly relativistic jets, such as unusually high-brightness temperatures requiring Doppler boosting \citep{Kellermann_Pauliny-Toth1969,Readhead1994}, for the radio-quiet NLS1s.  We will present a comprehensive study of the core activity of both radio-quiet and radio-loud NLS1s in another paper (Doi et~al.~in preparation).

\subsection{Extended Emission Components}\label{section:diffuseemission}
Past VLA surveys of NLS1s found a combination of two radio properties: compact sources ($\la300$~pc) and very steep spectra \citep{Ulvestad_etal.1995,Moran2000,Zhou_Wang2002}.  This had indicated that the radio sources should be localized, but physically not so compact that they become optically thin in synchrotron radiation.  Our VLBA observations have confirmed this picture at the pc scale.  

As an example, MRK~1239 is only slightly resolved in VLA images at arcsecond resolutions and is revealed to exhibit very diffuse structures within 40~pc in the VLBA image.  Less than half of the total fluxes were detected in the VLBA image and originate in low-brightness components of $\sim5\times10^6$~K (Appendix~\ref{section:individualobject}).  
We examine the possibility of the thermal process (free--free emission) for radio-emitting plasma of 15.2~mJy (the VLBA flux density of the extended component for MRK~1239).  The brightness temperature of $5\times10^6$~K gives a physical temperature of $\leq 5 \times10^6$~K.  According to the Gaunt factors of $\sim10$ and $\sim0.4$ at $\sim1$~GHz~\citep{Gayet1970} and $\sim1$~keV~\citep{Sutherland1998} for plasma at $\sim10^6$~K, the expected X-ray luminosity would be $>4\times10^{44}$~erg~s$^{-1}$ at 0.1--2.4~keV \citep{Rybicki_Lightman1979}, much higher than $1\times10^{42}$~erg~s$^{-1}$ observed by {\it ROSAT} toward MRK~1239 (\citealt{Zimmermann_etal.2001}, see a similar discussion for radio-quiet quasars by \citealt{Laor_Behar2008,Steenbrugge_etal.2011}).  It is difficult to reconcile a thermal process with the observed low-brightness radio emissions.  
The remaining lower brightness components ($\la10^6$~K) that were below the VLBA sensitivity are presumably nonthermal jets as well because the position angle of the VLA-resolved structure of MRK~1239 [Figure~\ref{figureA} (b) in Appendix~\ref{section:individualobject}] is consistent with the direction of the two-sided jet-like structure in the VLBA image.  
A steep spectrum in total flux [Figure~\ref{figureA} (c) Appendix~\ref{section:individualobject}] also strongly supports a nonthermal synchrotron origin for the diffuse emissions.  

MRK~705 and MRK~766 also exhibit clear morphological evidences of jets in the VLBA images.  A large portion of the total flux density comes from diffuse components ($f_\mathrm{1.7GHz}^\mathrm{10pc}=0.24$ and 0.07, respectively), which is also the case for MRK~1239 ($f_\mathrm{1.7GHz}^\mathrm{10pc}=0.23$).  
The other NLS1s also showed low core fractions ($f_\mathrm{1.7GHz}^\mathrm{10pc}\la0.10$).  
NGC~4051 was not detected in our observation with $f_\mathrm{1.7GHz}^\mathrm{10pc}<0.05$ but was detected by \citet{Giroletti_Panessa2009} at 1.6~GHz by very sensitive VLBI imaging.  The radio morphology is symmetric and shows two-sided lobe-like emissions separated by 20~pc and centered at a core of sub-mJy, which is aligned with larger-scale emissions in the VLA images (Appendix~\ref{section:individualobject}). 
There is an additional radio-quiet NLS1 that was previously observed by VLBI.  
The radio core of NGC~5506, a hidden NLS1 candidate \citep{Nagar_etal.2002a}, is surrounded by several diffuse components in the pc-scale region; a bridging morphology indicates a physical connection between the core and the diffuse components via jets \citep{Middelberg_etal.2004}.  Furthermore, a low-surface-brightness halo $\sim350$~pc in diameter enshrouds the central features in a VLA image \citep{Orienti_Prieto2010}, and only one-third of the single-dish flux was detected by VLBI \citep{Middelberg_etal.2004}.   
Thus, {\it a large portion of the radio luminosities come from diffuse emission components distributed within the nuclear regions ($\la300$~pc), which is a general property of non-thermal jets in radio-quiet NLS1s}.

These diffuse components are putative reservoirs of dissipated jet kinetic powers.  
For our samples ($\mathrm{M_{BH}} \sim 10^6$--$10^7$~M$_\sun$: Table~\ref{table1}) and the majority of nearby NLS1 populations \citep[$\la 10^7$~M$_\sun$; e.g.,][]{Zhou_etal.2006,Doi_etal.2012}, 
the jet kinetic powers limited by the Eddington limit on the low-mass black holes may be insufficient to cause the jets to escape to kpc scales outside the host galaxies.   
Indeed, only a small number of known NLS1s exhibit kpc-scale radio morphology \citep{Anton_etal.2008,Gliozzi_etal.2010,Doi_etal.2012}; all of these NLS1s possess relatively high-mass black holes of $\ga10^7$~M$_\sun$ and estimated jet kinetic powers of $\ga10^{44}$~erg~s$^{-1}$, and show signatures of relativistic jets in most cases \citep{Doi_etal.2012}.  
On the other hand, even the most luminous radio source in our radio-quiet sample (MRK~705) possesses an extended radio power of $10^{23.5}$~W~Hz$^{-1}$ at 1.4~GHz, suggesting an estimated jet kinetic power of $10^{43.5}$~erg~s$^{-1}$ \citep[by using the relation of][]{OSullivan_etal.2011}.  These values are below the thresholds of the Fanaroff--Riley type~I/II \citep{Fanaroff_Riley1974} dichotomy at a radio power of $\sim10^{24.6}$~W~Hz$^{-1}$ \citep{Zamfir_etal.2008} and a kinetic power of $\sim10^{44}$~erg~s$^{-1}$ \citep{Punsly_Zhang2011} that enables escape in the form of supersonic radio lobes (\citealt{Kawakatu_etal.2009}; see the discussion in \citealt{Doi_etal.2012} for NLS1s with kpc-scale radio structures).   
Furthermore, advancing radio lobes would be more likely to become disrupted by the dense interstellar medium in the nuclear region \citep[e.g.,][]{DeYoung1993} of spiral hosts, as seen in normal Seyfert galaxies \citep{Orienti_Prieto2010}.  
The diffuse components found in our NLS1 sample may also be relics of such disrupted structures.   
On the basis of the VLBI-observed radio-quiet and radio-loud NLS1s, we will present a study of their jet kinetic powers estimated from the extended radio components (Doi~et~al.~in~preparation).

\section{Summary and Conclusions}\label{section:summary}

Our VLBA imaging study at 1.7~GHz and mas resolution provide the first insight into the pc-scale region of several nearby radio-quiet NLS1s in a radio band.    
Five of seven sources (MRK~110, MRK~705, MRK~1239, MRK~766, and MRK~783) were detected at brightness temperatures of $\ga 5 \times 10^6$~K and contain radio cores with high-brightness temperatures of $>6 \times 10^7$~K, indicating a nonthermal process driven by a jet-producing central engine, in common with other AGN classes.      
MRK~1239 exhibits a two-sided jet-like structure $\sim40$~pc in extent.  
MRK~705 and MRK~766 exhibit one-sided linear jet structures of $\sim26$~pc and $\sim13$~pc, respectively.    
On the other hand, two sources (NGC~4051 and AKN~564) were not detected at the sensitivity of our observations.  

We defined the radio core fraction, $f_\mathrm{1.7GHz}^\mathrm{10pc}$, as the ratio of the flux density from the central 10~pc (in diameter) to the total flux density at 1.7~GHz in the rest frame.   
We found typical values of $f_\mathrm{1.7GHz}^\mathrm{10pc}<0.1$ and a maximum of 0.24 in our sample.    
While radio sources are apparently compact at arcsec resolution ($\la300$~pc), our VLBI study indicated that a large portion of the radio luminosities come from diffuse emission components that are distributed within the nuclear regions.  
Their central engines with low-mass black holes of $\sim10^6$--$10^7$~M$_{\sun}$ may supply jets with insufficient kinetic powers to escape to the outsides of the host galaxies.

\acknowledgments
We are grateful to the anonymous referee for offering constructive comments that have contributed in substantially improving this paper.  
The VLBA and VLA are operated by the NRAO, a facility of the National Science Foundation operated under cooperative agreement by Associated Universities, Inc.  
MERLIN is a National Facility operated by the University of Manchester at Jodrell Bank Observatory on behalf of STFC.  
We used NASA's Astrophysics Data System Abstract Service~(ADS) and the NASA/IPAC Extragalactic Database~(NED), which are operated by the Jet Propulsion Laboratory.  
This work was partially supported by a Grant-in-Aid for Scientific Research~(C; 21540250 and B; 24340042, AD) from the Japan Society for the Promotion of Science~(JSPS).  This work was also partially supported by the Center for the Promotion of Integrated Sciences~(CPIS) of Sokendai.

\appendix

\section{Individual radio-quiet NLS1s}\label{section:individualobject}

\subsection{MRK~110} 
Our VLBA image at 1.7~GHz (Figure~\ref{figure1}) showed a single component of 1.1~mJy, which retrieved 13\% of the total flux density in the rest frame on the basis of the 1.4-GHz NVSS flux density and a spectral index of $\alpha=-0.7$ between 1.4 and 8.4~GHz \citep{Kukula_etal.1998}.  Because we could not make self-calibration on this source owing to insufficient signal-to-noise ratios, we cannot rule out the possibility that the VLBA image was affected by some coherence loss in phase referencing.  VLA A-array images showed a secondary component 3\farcs0~(2.1~kpc) to the north of a central component (\citealt{Kukula_etal.1998}, see also \citealt{Miller_etal.1993}).

\subsection{MRK~705}
The self-calibrated VLBA image of MRK~705 at 1.7~GHz showed a compact core of 1.8~mJy, and a possible linear structure that extends $\sim45$~mas~($\sim26$~pc) to the east (Figures~\ref{figure1} and \ref{figure2}).  An unresolved source of 8.5~mJy was found in the FIRST image and accompanies a secondary source of 27.5~mJy, which has an unrelated infrared counterpart \citep{2005A_A...442..909G}.  The VLA A-array image at 8.46~GHz showed an unresolved ($<55$~pc) source of 2.0~mJy \citep{Schmitt_etal.2001}, which is consistent with the extent of the VLBA structure.  A single component of 5.7~mJy~beam$^{-1}$ was seen at 1.7~GHz in an archival image from the Multi-Element Radio Linked Interferometer Network~(MERLIN) at 150~mas resolution (MERLIN proposal code: Y/32); a large portion of the MERLIN fluxes was not retrieved in the VLBA baselines.  Because the emissions of MRK~705 were detected after self-calibration~(Section~\ref{section:datareduction}), its astrometric position could not be obtained.

\subsection{MRK~1239}
\begin{figure*}[h]
\epsscale{1.0}
\plotone{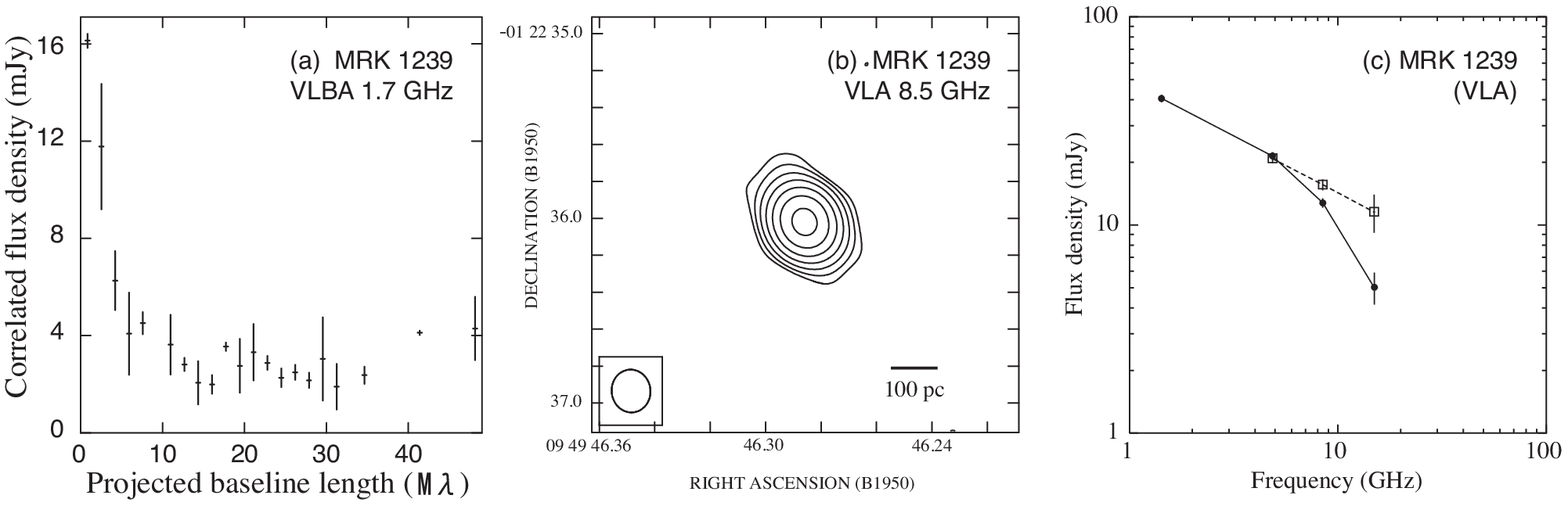}
\figcaption{(a)~Plot of visibility amplitude of MRK~1239 from VLBA data.  Visibilities integrated every 60~min are divided into 30~bins over the range of the baseline lengths.  (b)~VLA images of MRK~1239 with VLA A-array at 8.5~GHz.  Contour levels are $3\sigma \times -1$, 1, 2, 4, 8, 16, 32, and 64, where $\sigma=0.16$~mJy~beam$^{-1}$ is the image rms noise.  The FWHM of the synthesized beam, $0\farcs27 \times 0\farcs22$~at $\mathrm{PA}=7.6$\degr, is illustrated in the lower left corner.  (c)~Radio spectra of MRK~1239 with VLA B-array.  Filled circles connected with solid lines represent measurements from VLA images with uniform weighting.  Open squares connected with dashed lines represent measurements from tapered images with a common beam size using the same data.\label{figureA}}
\end{figure*}

MRK~1239 is an unresolved ($<80$~pc) source showing a steep spectrum in VLA A-array images \citep{Ulvestad_etal.1995,Thean_etal.2000}.  However, our VLA A-array image from a good $uv$-coverage at 8.5~GHz from the archival data AM384 showed a significantly elongated structure of $\sim0\farcs12$, corresponding to $\sim46$~pc, at a position angle of 47\degr~(Figure~\ref{figureA} (b)).  

The self-calibrated VLBA image of MRK~1239 at 1.7~GHz shows a low-brightness (2.4~mJy beam$^{-1}$ or less) linear structure of $\sim100$~mas ($\sim40$~pc) consisting of several components~(Figure~\ref{figure1}), which is consistent with the elongated size seen in the VLA image.  The visibility amplitudes were clearly enhanced in baselines at $\la3$~M$\lambda$~[Figure~\ref{figureA} (a)].  A visibility-based model fitting with a single-Gaussian source structure resulted in a major axis of 62~mas ($\sim$24~pc) at a position angle of 47$\degr$.  The flux density of the Gaussian model was 21.5~mJy, consistent with the integrated flux density of 20.0~mJy in the VLBA image~(Table~\ref{table3}).  \citet{Orienti_Prieto2010} also reported a VLBA image of MRK~1239 at 1.7~GHz that showed two components of 10.2~mJy in total aligned at a position angle similar to our result.  The contribution from the central 10~pc was 23\% of the total flux density at the rest frame on the basis of the 1.4-GHz NVSS flux density and assumed spectral indices of $\alpha=-0.5$ for the core and $\alpha=-0.54$ (below) for the extended components, respectively.  

The presence of diffuse emission distributed within $\sim50$~pc can also be inferred from our multi-frequency VLA B-array (quasi-simultaneous) observations~[AD489; Figure~\ref{figureA} (c)].  A spectral index of $\alpha=-0.52$ was found at 1.4--5~GHz; steeper spectra appear at higher frequencies, $\alpha=-0.94$ at 5--8.5~GHz and $\alpha=-1.64$ at 8.5--15~GHz ($S_\nu \propto \nu^{\alpha}$).  Such an apparent spectral break also appeared in the VLA A-array (not simultaneous) observations at 1.4--8.4~GHz (not simultaneously) by \citet{Ulvestad_etal.1995}.  \citet{Moran2000} pointed out that NLS1s tend to have significantly steeper radio spectra ($\alpha\sim1.1$--1.2) than classical Seyferts on the basis of VLA A-configuration observations at 1.4 and 8.4~GHz.  They noted that the strong radiation field around highly accreting disks of NLS1s can steepen the synchrotron spectra owing to inverse-Compton cooling of relativistic electrons in jets.  However, our measurements for MRK~1239 from $uv$-tapered images of a common resolution resulted in a simple power-law spectrum with $\alpha=-0.54$.  This provides evidence for resolution effects.  The significantly steeper spectra generally seen in radio-quiet NLS1s might have been artificial, which may be related to the dominance of diffuse components (Section~\ref{section:diffuseemission}).

\subsection{NGC~4051}\label{section:NGC4051} 
NGC~4051 is one of two NLS1s not detected in our VLBA observation.  To determine the phase-tracking center, we reanalyzed the VLA A-array 22-GHz image of \citet{Hagiwara_etal.2003}, which included nuclear H$_2$O masers confined to a region of 0\farcs04; the VLA observation used the ICRF calibrator J1153+4931 as a reference position.  \citet{Giroletti_Panessa2009} reported VLBI detection of NGC~4051 using the European VLBI Network~(EVN) at both 1.6 and 5~GHz at a position that differed by only $<10$~mas from our phase-tracking center (Table~\ref{table2}).  Hence, it was unlikely that the nucleus of NGC~4051 was outside of the VLBA's FOV.

The EVN image at 1.7~GHz showed a core (0.45~mJy) and two symmetric sub-components (0.73 and 0.67~mJy) separated by $\sim20$~pc with respect to the core with a maximum intensity of $\sim0.34$~mJy~beam$^{-1}$ \citep{Giroletti_Panessa2009}.  Such weak emissions are consistent with our negative detection because they are  $3\sigma$ in the VLBA image.  A MERLIN image at 1.6~GHz showed a triple structure with an extent of 0\farcs8 (38~pc) at a position angle of 73\degr~\citep{Christopoulou_etal.1997}, which corresponds to the structure seen in the EVN image.  The VLA A-array image at 1.4~GHz showed larger structure extending up to $\sim4$\arcsec~($\sim330$~pc) from the nucleus \citep{Ho_Ulvestad2001,Baum_etal.1993}; the position angle has changed to 41\degr, perpendicular to the host-galaxy major axis (135\degr).  The NVSS image shows very diffuse emission associated with the spiral arms and disk at a scale of arcmin, indicating stellar origins \citep{Baum_etal.1993}.  Hence, we used a flux density of 14.7~mJy in the VLA A-array image at 1.4~GHz presented by \citet{Ho_Ulvestad2001} as the total flux density from an AGN contribution to derive the core fraction, assuming  $\alpha=-0.7$ \citep{Christopoulou_etal.1997}.    

\subsection{MRK~766} 
The self-calibrated VLBA image with natural weighting at 1.7~GHz showed marginal structures extending to the southwest (PA$\sim-142$\degr) and north (PA$\sim0$\degr) from the brightest component (Figure~\ref{figure1}).  In a restored image, diffuse emissions extending up to $\sim$50~mas (corresponding to $\sim$13~pc) north were clearly seen (Figure~\ref{figure2}).  We recognized an enhancement in the visibility amplitude of MRK~766 at baselines of $<5$~M$\lambda$; this suggests the presence of a significant amount of diffuse component with the size of $>10$~pc.  Earlier VLA A-array images showed a slightly resolved structure ($\la100$~pc) at a position angle of $\sim30$\degr\ \citep{Ulvestad_etal.1995,Nagar_etal.1999,Thean_etal.2000}.  A MERLIN image at 1.7~GHz clearly showed a radio structure extending to 0\farcs25~(65~pc) at a position angle of 34\degr~\citep{Thean_etal.2001}, which is consistent with the southwest elongation in our VLBA image with natural weighting.  The VLBA flux density was 2.5~mJy, which retrieved 7\% of the total flux density at the rest frame on the basis of the 1.4-GHz NVSS flux density and $\alpha=-0.72$ between 1.4 and 8.4~GHz \citep{Thean_etal.2001}.  \citet{Lal_etal.2004} reported a VLBI detection of MRK~766 at a relatively high intensity 7.1~mJy~beam$^{-1}$ with a synthesized beam of $\sim$2~mas at 5~GHz, which was inconsistent with a MERLIN archival image at 5~GHz (MERLIN proposal code: 92NOVA) that showed diffuse emission of only 1.8~mJy~beam$^{-1}$; this discrepancy might be because of variability.  Because of the negative detection before self-calibration~(Section~\ref{section:VLBAobservation}), the astrometric position could not be obtained from our VLBA observation.

\subsection{MRK~783}
Our VLBA image at 1.7~GHz (Figure~\ref{figure1}) showed a single component of 1.3~mJy, which retrieved 4\% of the total flux density at the rest frame on the basis of the 1.4-GHz NVSS flux density and assumed spectral indices of $\alpha = -0.5$ and $-0.7$ for the VLBA and diffuse components, respectively.  The VLA A-array image at 4.9~GHz showed an unresolved ($<380$~pc) source of 3.5~mJy \citep{Ulvestad_Wilson1984}.

\subsection{AKN~564}
AKN~564 is one of two NLS1s not detected in our VLBA observation.  The VLA A-array image at 8.46~GHz showed a triple structure (3.1, 3.0, and 0.9~mJy) extending 320~pc to the north \citep{Schmitt_etal.2001,Moran2000}.  A MERLIN archival image showed somewhat diffuse emissions of only 1.04~mJy~beam$^{-1}$ at 5~GHz with a 40~mas beam, which seemed to be consistent with our negative detection.  On the other hand, a mas-scale structure with 3.7~mJy~beam$^{-1}$ had been detected at 5~GHz by conventional fringe fitting~(i.e., not phase referencing) using a much more sensitive VLBI array of $\sim$2~mas resolution at 5~GHz by \citet{Lal_etal.2004}.  However they mentioned that they were not confident about the structure.  THe phase-tracking center of our VLBA observation was determined from the VLA archival data AB973 obtained with the A-array at 8.4~GHz (Table~\ref{table1}) using the ICRF position of the phase calibrator J2236+2828 and differed by only $\sim$260~mas from that of \citet{Lal_etal.2004}.  Hence, it was unlikely that the nucleus of AKN~564 was outside of the VLBA's FOV.  Some amount of diffuse emissions could be in the VLBA FOV because the correlated amplitude was enhanced at baselines of $\la5$~M$\lambda$, and the image noise was significantly higher than the expected thermal noise level~(Section~\ref{section:datareduction}).  Hence, we concluded that AKN~564 is dominated by diffuse components at pc scales, which are responsible for our negative detection.  The spectral index between 1.4 and 4.9~GHz is $\alpha=-0.74$ \citep{Ulvestad_etal.1981}.

\end{document}